\documentclass[11pt]{article}
\title{\bf Fluctuation Theorem and Chaos}
\author{\small\textsc{Giovanni Gallavotti}%
\\
\small Dipartimento di Fisica and INFN\\
\small Universit\`a di Roma 
{\it La Sapienza}\\
\small P.~A.~Moro 2, 00185, Roma, Italy\\
\small \texttt{giovanni.gallavotti@roma1.infn.it}
}
\date{\today}
\newcount\biblio%
\font\ottorm=cmr8%
\font\diecibf=cmbx10%
\font\cs=cmcsc10\font\sc=cmcsc10%
\font\ottorm=cmr8%
\font\tenmib=cmmib10 \font\eightmib=cmmib8
\font\sevenmib=cmmib7\font\fivemib=cmmib5 
\font\ottoit=cmti8\font\fiveit=cmti5\font\sixit=cmti6%%
\font\fivei=cmmi5\font\sixi=cmmi6\font\ottoi=cmmi8
\font\ottorm=cmr8
\font\ottosy=cmsy8\font\sixsy=cmsy6\font\fivesy=cmsy5%%
\font\ottocss=cmcsc8%
\def\ottopunti{\def\rm{\fam0\ottorm}\def\it{\fam6\ottoit}%
\textfont1=\ottoi\scriptfont1=\sixi\scriptscriptfont1=\fivei%
\textfont2=\ottosy\scriptfont2=\sixsy\scriptscriptfont2=\fivesy%
\textfont3=\tenex\scriptfont3=\tenex\scriptscriptfont3=\tenex%
\textfont4=\ottocss\scriptfont4=\sc\scriptscriptfont4=\sc%
\textfont5=\eightmib\scriptfont5=\sevenmib\scriptscriptfont5=\fivemib%
\textfont6=\ottoit\scriptfont6=\sixit\scriptscriptfont6=\fiveit%
\textfont\bffam=\eightmib\scriptfont\bffam=\sevenmib%
\scriptscriptfont\bffam=\fivemib%
\setbox\strutbox=\hbox{\vrule height7pt depth2pt width0pt}%
\normalbaselineskip=9pt\rm}
\mathchardef\Ba   = "050B  %alfa
\mathchardef\Bb   = "050C  %beta
\mathchardef\Bg   = "050D  %gamma
\mathchardef\Bd   = "050E  %delta
\mathchardef\Be   = "0522  %varepsilon
\mathchardef\Bee  = "050F  %epsilon
\mathchardef\Bz   = "0510  %zeta
\mathchardef\Bh   = "0511  %eta
\mathchardef\Bthh = "0512  %teta
\mathchardef\Bth  = "0523  %varteta
\mathchardef\Bi   = "0513  %iota
\mathchardef\Bk   = "0514  %kappa
\mathchardef\Bl   = "0515  %lambda
\mathchardef\Bm   = "0516  %mu
\mathchardef\Bn   = "0517  %nu
\mathchardef\Bx   = "0518  %xi
\mathchardef\Bom  = "0530  %omi
\mathchardef\Bp   = "0519  %pi
\mathchardef\Br   = "0525  %ro
\mathchardef\Bro  = "051A  %varrho
\mathchardef\Bs   = "051B  %sigma
\mathchardef\Bsi  = "0526  %varsigma
\mathchardef\Bt   = "051C  %tau
\mathchardef\Bu   = "051D  %upsilon
\mathchardef\Bf   = "0527  %phi
\mathchardef\Bff  = "051E  %varphi
\mathchardef\Bch  = "051F  %chi
\mathchardef\Bps  = "0520  %psi
\mathchardef\Bo   = "0521  %omega
\mathchardef\Bome = "0524  %varomega
\mathchardef\BG   = "0500  %Gamma
\mathchardef\BD   = "0501  %Delta
\mathchardef\BTh  = "0502  %Theta
\mathchardef\BL   = "0503  %Lambda
\mathchardef\BX   = "0504  %Xi
\mathchardef\BP   = "0505  %Pi
\mathchardef\BS   = "0506  %Sigma
\mathchardef\BU   = "0507  %Upsilon
\mathchardef\BF   = "0508  %Fi
\mathchardef\BPs  = "0509  %Psi
\mathchardef\BO   = "050A  %Omega
\mathchardef\BDpr = "0540  %Dpr
\mathchardef\Bstl = "053F  %*

\let\a=\alpha      \let\e=\varepsilon
\let\z=\zeta       
\let\m=\mu                 
\let\s=\sigma \let\t=\tau   \let\f=\varphi 
   
\let\G=\Gamma \let\D=\Delta   
    \let\Si=\Sigma     \let\Ps=\Psi

\def\\{\hfill\break} \let\==\equiv

\let\io=\infty 

\let\0=\noindent

\def\media#1{{\langle#1\rangle}}
\def\ie{\hbox{\it i.e.\ }}
\let\dpr=\partial  

\def\tende#1{\,\vtop{\ialign{##\crcr\rightarrowfill\crcr
 \noalign{\kern-1pt\nointerlineskip} \hskip3.pt${\scriptstyle
 #1}$\hskip3.pt\crcr}}\,}
\def\circage{\lower2pt\hbox{$\,\buildrel > \over {\scriptstyle \sim}\,$}}
\def\otto{\,{\kern-1.truept\leftarrow\kern-5.truept\to\kern-1.truept}\,}
\def\fra#1#2{{#1\over#2}}
\def\CC{{\cal C}}

\def\T#1{{#1_{\kern-3pt\lower7pt\hbox{$\widetilde{}$}}\kern3pt}}
\def\VVV#1{{\underline #1}_{\kern-3pt%
\lower7pt\hbox{$\widetilde{}$}}\kern3pt\,}
\def\W#1{#1_{\kern-3pt\lower7.5pt\hbox{$\widetilde{}$}}\kern2pt\,}
\def\wt{\widetilde}

\def\lis{\overline}

\def\indica{\leaders \hbox to 0.5cm{\hss.\hss}\hfill}
\def\guida{\leaders\hbox to 1em{\hss.\hss}\hfill}
\def\qed{\raise1pt\hbox{\vrule height5pt width5pt depth0pt}}

\def\V#1{{\bf#1}}
\def\defi{\,{\buildrel def\over=}\,}

\def\sqr#1#2{{\vcenter{\vbox{\hrule height.#2pt%
	\hbox{\vrule width.#2pt height#1pt \kern#1pt%
	  \vrule width.#2pt}%
	\hrule height.#2pt}}}}

\def\ig{\int}

\def\*{\vglue0.3truecm}
\renewcommand{\theequation}{\arabic{section}.\arabic{equation}}
\def\be{\begin{equation}}
\def\ee{\end{equation}}

\newdimen\xshift \newdimen\xwidth \newdimen\yshift \newdimen\ywidth

\def\ins#1#2#3{\vbox to0pt{\kern-#2\hbox{\kern#1 #3}\vss}\nointerlineskip}

\def\eqfig#1#2#3#4#5{
\par\xwidth=#1pt \xshift=\hsize \advance\xshift
by-\xwidth \divide\xshift by 2
\yshift=#2pt \divide\yshift by 2
{\hglue\xshift \vbox to #2pt{\vfil
#3 \includegraphics{#4.eps}
}\hfill\raise\yshift\hbox{#5}}}

\def\8{\write12}  %abbreviazione{\openout15=\jobname.aux}

\usepackage{makeidx}         % allows index generation
\usepackage{eqalignno}

\usepackage{fancyhdr}
\font\diecibf=cmbx10%
\renewcommand{\headrulewidth}{0pt}
\def\iniz{\setcounter{equation}{0}{
\rhead{\thepage}\lhead{{{{\diecibf\thesection:}\ \SEC}}}}}

\begin{document}
\maketitle
\begin{abstract}{\it The heat theorem  (i.e. the second law of
    thermodynamics or the existence of entropy) is a manifestation of
  a general property of hamiltonian mechanics and of the ergodic
  Hypothesis. In nonequilibrium thermodynamics of stationary states
  the chaotic hypothesis plays a similar role: it allows a unique
  determination of the probability distribution (called {\rm SRB}
  distribution on phase space providing the time averages of the
  observables. It also implies an expression for a few averages
  concrete enough to derive consequences of symmetry properties like
  the fluctuation theorem or to formulate a theory of coarse graining
  unifying the foundations of equilibrium and of nonequilibrium.
  }\end{abstract} \*
%$\Bf=\Ba=\Bb$.
%\vfill\eject

\def\SEC{Boltzmann's heat theorem}
\section{Boltzmann's Heat Theorem}\iniz
\*

In equilibrium statistical mechanics states are identificed with time
invariant probability distributions $\m$ on phase space. Thermodynamic
functions, identified with time averages of mechanical observables,
are expressed as integrals $\media{F}$ of suitable mechanical
obeservables $F$. The averages depend on control parameters $\Ba$,
like volume, energy, kinetic energy. Under changes b $d\Ba$ of the
control parameters the thermodynamic quantities change so that the
variation of the average energy and the variation of the volume are
$dU$ and $dV$ and are related to the time averages of the kinetic
energy $\media{K}\defi\media{\sum_{i=1}^N \frac{m\dot{\V x}_i^2}2}
\defi\frac32 N k_B T$ and of $p\defi\media{-\dpr_V U}$, with $U$ being
the total potential energy, so that, expressing $p,T$ in terms of the
control parameters:
\*

\0{\bf Heat Theorem:} (HT) {\it Changing $\Ba\to\Ba+d\Ba$ induce
changes $dU$, $dV$, with

\be\fra{dU+p dV}T= {\rm exact}\defi dS,\label{e1.1}\ee
under the ergodic hypothesis, or equivalently under the assumption
that the distributions $\m$ are elements of one among the classical
ensembles, like microcanonical, canonical,$\ldots$,\rm
\cite{Bo866,Bo884}.}

\* In modern terminology: the {\it ergodic hypothesis} (EH) implies
equilibrium statistical mechanics. The guiding idea is that HT holds
for {\it all} (ergodic) systems with Hamiltonian of the form $H=K+U$:
{\it whether having few {\rm($\sim1$)} or many {\rm ($\sim10^{19}$)}
degrees of freedom}, as long as EH holds. This means that HT is a
trivial consequence of the Hamiltonian structure of the mechanical
systems describing the microscopic motions. It is always valid, like a
{\it symmetry property}, and it is highly nontrivial in systems with
many degrees of freedom, being the second law of thermodynamics.  In
other words a guiding idea to understand certain universal laws is
that they merely reflect symmetries or general stuctures, of the
underlying equations, which may have deep consequences in
large systems: e.g, via the HT, the roots of second law can
be found, \cite{Bo866}, in the simple properties of the pendulum.

\*

\def\SEC{Thermostats and reversiblity}
\section{Thermostats and reversiblity}\iniz

Stationary states out of equilibrium are realized when on a system are
present staionary currents. In such systems currents generate, by
dissipation, heat that is absorbed by thermostats.

Recent progress has been achieved by employing simple models of the
thermostats with the feature of being {\it finite} systems of
particles, hence well suited for simulations. There are various
types of thermostats considered in the literature.  As a rather
general class of thermostats model consider

\eqfig{110}{80}{}{fig1}{}

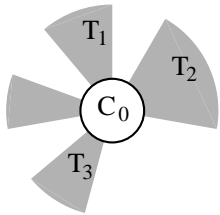
\0Fig.1: {\small $\CC_0$ (``system'') interacts with shaded $T_j$
(``thermostats'') constrained to keep fixed kinetic energy
$K_j=\frac{m}2\dot{\V X}^2_j=\frac32 N_j k_B T_j$.}  \*

The equations of motion for the $N_0,N_1,\ldots$ particles located in
configurations $\V X=(\V X_0,\V X_1,\ldots)$ inside the containers
$\CC_0,T_1,\ldots$ (if here $\V X_j=(\V x_{j,1},\ldots, \V x_{j,N_j}$)
will be (as an example with all masses equal), $\V
E$= positional ``stirring forces'', $U_j$ internal energy in the
$j$-th container, $W_{0,j}$ potential energy of interaction between
the particles in $\CC_0$ and those in $\CC_j$)

\be\eqalign{
m\ddot{\V X}_0=&-\dpr_{\V X_0}\Big( U_0(\V X_0)+\sum_{j>0}
W_{0,j}(\V X_{0},\V X_j)\Big)+\V E(\V X_0),
\cr
m\ddot{\V X}_i=&-\dpr_{\V X_i}\Big( U_i(\V X_i)+
W_{0,i}(\V X_{0},\V X_i)\Big)-\a_i \dot{\V X}_i\hbox{\hglue1.1truecm}
\cr}\label{e2.1}\ee
The energies $U_0,U_j,W_{0,j}$ should be imagined as generated by pair
potentials $\f_0,\f_j,\f_{0,j}$ short ranged, smooth, or with a
singularity like Lennard-Jones type at contact, and by external
potentials modeling the containers walls), $\a_i$ determined so that
$K_i=\fra32 N_i k_B T_i\= \,const$.

More generally thermostats can even act on regions of $\CC_0$: eg. in
electric conduction models analogous to Drude's model (1899!), one
imagines that the collisions with the lattice communicate energy to the
lattice vibrations (``phonons'') and this is modeled by adding a
constraint that $\frac{m}2\dot{\V X}_0^2=\frac32k_B T_0$ keeping the
total kinetic energy of the $N_0$ particles in $\CC_0$ identically constant
realized by an extra term $-\a_0\dot{\V X}_0$ in the first of
Eq.(\ref{e2.2}) with $\a_0$ suitably chosen. The multipliers $\a_j$ in
Eq.(\ref{e2.2}) are readily computed by imposing constancy of 
$ K_i\=const\defi\fra32 N_i k_B T_i$ and are

\be
\a_i\=\fra{Q_i-\dot U_i}{3N_i k_B T_i}\label{e2.2}\ee
and $Q_i\=$ work per unit time done by $\CC_0$ on $\CC_i$:

\be Q_i\defi-\dot{\V
X}_i\cdot\dpr_{{\V X}_{i}}W_{0,i}(\V X_{0},{\V X}_i)\label{e2.3}\ee
and it is naturally  interpreted as the {\it heat} ceded per unit time
to the thermostat $\CC_i$.

The  main feature of the above equations is that they are {\it not}
Hamiltonian and, therefore, the phase space volume measured by the
divergence $\s(\V X,\dot{\V X})$ is not zero, and after an algebraic
computation is checked  to be (neglecting for simplicity
factors of the form $(1-\frac{1}{3 N_j})$)

\be\eqalign{
\s(\V X,\dot{\V X})&\=\e(\V X,\dot{\V X})+
\dot W(\V X),\cr
\e(\V X,\dot{\V X})&=\sum_{j=1}^n \fra{Q_j}{k_B T_j}, \quad
W=\sum_j\fra{U_j(\V X_j)}{k_B T_j}}\label{e2.4}\ee
This is a sum of two terms, one of which has the interpretation of
entropy increase of the thermostats per unit time while the other is a
time derivative. Therefore one term is accessible not only in
simulations but it is also conceivable that it can be measured in
experiments; the other is instead strongly model dependent,
coordinates dependent and metric dependent.  

Abridging often $(\V X,\dot{\V X})$ simply by $x$ and changing
coordinates or metric the expression for $\s$ changes as
$\s'(x)=\s(x)+\fra{d}{dt} \G(x)$ with a suitable $\G$.  Therefore {\it
only} time averages over long times can have ``intrinsic'' meaning
because

\be\fra1\t\ig_0^\t \s'(S_tx)dt=\fra1\t\ig_0^\t \s(S_tx)dt+
\fra{\G(S_\t x)-\G(x)}\t\label{e2.5}
\ee 
so that the two averages of $\s$ and $\s'$ have the same limiting
behavior as $\t\to\infty$, at least if $\G$ is bounded. Hence if the
$U_j$ are bounded (as we shall suppose for simplicity) and if the average
$\s_+\defi\mathop{\rm lim}\limits_{\t\to+\io}\fra1\t\ig_0^\t \s(S_tx)$ exists
then it can be identified to entropy creation rate

\be\s_+=\media{\sum_j\fra{Q_j}{k_BT_j}}.\label{e2.6}\ee
{\it Furthermore} the probability distributions in the stationary
states  of the averages of $\s$ and $\e$ over finite time $\t$
coincide asymptotically as $\t\to\infty$ because 

\be \fra1\t\ig_0^\t \s(S_t x)dt\=\fra1\t\ig_0^\t \e(S_t
x)dt+\fra{W(\t)-W(0)}\t:\label{e2.7}\ee
for large $\t$ the averages of $\s$ and $\e$ have the {\it same
fluctuations statistics}. Not just the same average:
$\media{\s}\=\media{\e}$. 

Hence {\it a general theory of fluctuations of long time averages of
$\s$, if at all possible, will imply a general theory of fluctuations
of $\e$}. And the latter is a quantity accessible experimentally via
calorimetric and thermometric measurements without need of the
equations of motion.

A further {\it important feature} of the model is that its equations,
Eq.)\ref{e2.1}), have a {\it time reversal symmetry}.  This means that
there exists a map $I$ of phase space which is isometric and smooth
with $I^2=1$ and $IS_t=S_{-t}I$ if $t\to S_t(\V X,\dot{\V X})$ denotes
the solution to the equations of motion with initial datum $(\V
X,\dot{\V X})$, $\V X\defi(\V X_0,\V X_1,\ldots)$. In this case $I$
can be simply defined as $I(\V X,\dot{\V X})\defi (\V X,-\dot{\V X})$.
\vfill\eject

\def\SEC{Chaotic hypothesis}
\section{Chaotic hypothesis}\iniz

Having identified entropy creation rate with a microscopic mechanical
quantity has been a key step towards the understanding of
nonequilibrium steady states. In a way it might turn out to be as
important as the realization, marking the beginning of statistical
mechanics, that in equilibrium the average kinetic energy has to be
identified with the absolute temperature.

To turn the above ``discovery'', \cite{EM90,Ga06c}, into a few quantitative
predictions of properties of steady nonequilibrium states it is
necessary to identify the probability distributions on phase space
that can be used to yield the ime averages of the observables.

The difficulty is that unless $\s_+=0$, Eq.(\ref{e2.6}), such
probability distributions must give probability $1$ to a set of data
which has $0$ volume.  This kind of problem arose in the theory of
turbulence and was solved by Ruelle's proposal,
\cite{Ru78b}, that the system (for
instance the Navier Stokes evolution) is so chaotic that it can be
regarded as having an axiom A attractor.

The idea has been extended to the dynamics of thermostatted systems.
It is convenient to formulate it in terms of a map $S$ between {\it
timing events}, i.e. by imagining to perform observations every time a
prefixed event takes place or, mathematically, every time the trajectory
crosses a prefixed surface $\Si$ in phase space. The time evolution
can then be described by a map $S$ defined on $\Si$ and mapping an
$x\in\Si$ into the next point $Sx$ where the trajectory of $x$ crosses
again $\Si$ (``{\it Poincar\'e's map} on $\Si$''). In this case the
phase space contraction is the logarithm of the Jacobian determinant
of the map $S$: namely $\wt \s(x)\defi-\log|\det \dpr S(x)|$. 

It should be remarked that the time between two successive events is
in general variable as a function of the point $x$: calling it $t(x)$
the map $S$ and the solutions $t\to S_tx$ of the equations of motion
are related by $Sx\equiv S_{t(x)}x$ and therefore for $x\in\Si$
it is $\wt \s(x)=\ig_0^{t(x)} \s(S_tx)dt$, and $\ig_0^\t F(S_tx)dt=
\sum_{k=0}^N \wt F(S^kx)$. 

The mentioned extension is obtained by formulating the \*

\0{\bf Chaotic hypothesis} {\it {\rm(CH)} Motions developing on the
attracting set for map $S$ representing the evolution of a chao\-tic
system of particles, observed in discrete time via a choice of timing
events $\Si$, may be regarded as motions of transitive hyperbolic
system. } \*

Informally such a system (also called {\it Anosov system}) has a
dynamics with the property that following the motion of any initial
datum $x$ the nearby points separate from it exponentially fast, in
the future {\it and} in the past, except when located on a surface
$W_s(x)$ through $x$ or, respectively, on another surface $W_u(x)$.

The assumption has to be understood in the same sense as the EH: the
 latter, as in {\cs Ruelle}'s view in \cite{Ru73}, can be commented as
``{\it ... while one would be very happy to prove ergodicity because it
would justify the use of Gibbs' microcanonical xensemble, real systems
perhaps are not ergodic but behave nevertheless in much the same way
and are well described by Gibbs' ensemble...}''.

Under the CH the following properties hold:
\*
\0(1) there is a {\it unique} distribution $\m$ such that,
for all $x$ outside a set of {\it zero volume},

\be\lim_{N\to\io}\fra1N \sum_{k=0}^{N-1}F(S^kx)=\ig \m(dy) F(y).\label{e3.1}\ee

\0(2) the probability distribution $\m$ has an ``explicit''
expression ``similar to the equilibrium Gibbs distribution'', \cite{Ga08}.

\0(3) $\m$ is concentrated on a $0$ volume ``attractor''.
\*

The distribution $\m$ is called the ``SRB distribution'' (acronym for
Sinai-Ruelle-Bowen).  Because of the above properties Anosov
maps are considered a {\it paradigm of chaos}, much as harmonic
oscillators are considered paradigms of {\it order}.  They enjoy
several interesting properties. Consider the finite time average
$f\defi\fra1N\sum_{k=0}^{N-1} F(S^kx)$, then
\*

\0{\bf Fluctuation Law}: {\it There are values $f_1,f_2$ such that
$f$ is in $[a,b]\subset (f_1,f_2)$ with $\m$-probability
$Prob_\m(f\in[a,b])\sim e^{\t \z_F(f)}$ in the sense that

\be\lim_{N\to\io} 
\fra1N\log Prob_\m(f\in[a,b])= \max_{f\in[a,b]} \z_F(f),\label{e3.2}\ee
and $\z_F(f)$ is analytic and convex in $(f_1,f_2)$.  \\ More
generally,\cite{Ga97}, given $n$ observables $\V F=(F_1,\ldots,F_n)$,
there exists a convex open set $\G\subset R^m$, $m\le n$, with the
property that if $\D\subset \G $ is a closed set and $f_j\defi
\fra1N\sum_{k=0}^{N-1} F_j(S^k x)$, then

\be Prob_\m(\V f\in\D)=
Prob_\m((f_1,\ldots,f_n)\in\D)\propto_{\t\to\io} e^{\t\max_{\V
    f\in\D}\z(\V f)}
\label{e3.3}\ee
with $\z_{\V F}$ analytic and convex in $\G$.
} (Sinai, \cite{GBG04}).
\*

The function $\z_{\V F}$ is a kind of thermodynamic function and via
the mentioned expression of $\m$ it is possible to obtain an explicit
(generally ``uncomputable'') expression of stationary averages
$\media{F}_\m$ and of $\z_{\V F}$.

\*
\def\SEC{Fluctuation theorem}
\section{Fluctuation Theorem}\iniz

Consider time reversal symmetric evolutions, see Sec.2. If the dynamics is a
discrete one, associated with a Poincar\'e section $\Si$ and a time
reversal symmetric evolution, a time reversal symmetry $I$ will be
smooth map $I$ of $\Si$ with the properties $I^2=1$ and
$IS=S^{-1}I$. It can be obtained by restricting to the timing events
map the symmetry in continuous time provided the Poincar\'e section
$\Si$ is chosen so that $I\Si=\Si$ (just replace $\Si$ by $\Si\cup I
\Si$). 

Assume :\\
(1) {\it chaotic hypothesis}\\
(2) {\it dissipativity}, i.e. the average phase space contraction $\s_+$, in
Eq(\ref{e2.6}), is positive, $\s_+>0$, {\bf and} \\
(3) {\it time reversal symmetry} by a map $I$.

Let $F_1\=\fra{\s}{\s_+}$ and $p\defi f_1 =\fra1N\sum_{k=0}^{N-1}
\fra{\s(S^k x)}{\s_+}$ where $\s(x)\defi-\log|\det \dpr S)|$. Then,
\cite{GC95,Ga95b}, \*

\0{\bf Fluctuation Theorem (FT):} {\it There is $p^*\ge1$, see
\cite{Ga95b}, such that the symmetry

\be\z(-p)=\z(p)-p\s_+,\qquad |p|< p^*\label{e4.1}\ee
holds.  More generally if $F_2,\ldots,F_n$ are any other $n-1$
functions of well defined parity under time reversal (i.e. even,
$F_j(Ix)=F_j(x)$, or odd, $F_j(Ix)=-F_j(x)$) then setting $I F_j=F_j$
for $F_j$ even and $IF_j=-F_j$ for $F_j$ odd it is

\be\z(-p,If_2,\ldots,If_n)=\z(p,f_2,\ldots,f_n)-p\s_+\label{e4.2}\ee
}

The physical interpretation of $p \s_+$ as the average of the 
the thermostats entropy increase rate  $\e_N=
\fra1N\sum_{k=0}^{N-1} \fra{Q_j}{k_B T_j}$, makes the theorem of
physical interest because, as mentioned, $\e_N$ is a measurable quantity
{\it independently of the model}. 

The Eq.(\ref{e4.2}) is a special case of an even more general relation
that is closely related to the Onsager-Machlup theory of {\it
fluctuation patterns}, \cite{Ga99,Ga00,Ga02}.  The question is which
is the probability that the successive values of $F_j(S_t x)$ follow,
for $t\in[-\t,\t]$, a preassigned sequence of values, that will be
called {\it pattern} $\f(t)$, \cite{Ga97}.

In a reversible hyperbolic and transitive system consider $n$
observables $F_1,\ldots,F_n$ which have a well defined parity under
time reversal $F_j(Ix)=\pm F_j(x)$. Given $n$ functions $\f_j(t)$,
$j=1,\ldots,n$, defined for $t\in[-\fra\t2,\fra\t2]$ the question is:
which is the probability that $F_j(S_tx)\sim \f_j(t)$ for $t\in
[-\fra\t2,\fra\t2]$? The following,\cite{Ga97},
gives an answer: \*

\0{\bf Fluctuation Patterns Theorem (FPT):} {\it Under the assumptions
of the fluctuation theorem given $F_j,\f_j$, and given $\e>0$ and an
interval $\D\subset(-p^*,p^*)$ the joint probabilities with respect to
the SRB distribution that $F_j(S_tx) $ follows the pattern $\f_j(t)$ or the
``time reversed pattern'' $\pm\f_j(-t)$ (the sign depending on the
parity of $F_j$) are related by

\be\eqalign{
&\fra{P_\m(|F_j(S_tx)-\f_j(t)|_{j=1,\ldots,n}<\e, p\in\D)}
{P_\m(|F_j(S_tx)\mp\f_j(-t)|_{j=1,\ldots,n}<\e, -p\in\D)}=\cr
&=\exp(\t
  \max_{p\in\D}\,p\,\s_++O(1))\cr}\label{e4.3}\ee
where sign choice $\mp$ is opposite to the parity of $F_j$ and
$p\defi \fra1\t\ig_{-\fra\t2}^{\fra\t2} \fra{\s(S_tx)}{\s_+}\,dt$.
The relation holds for patterns which can be
realized with a probability that does not vanish faster than
exponentially in time.}
\*

The FPT theorem means that ``all that has to be
done to change the time arrow is to change the sign of the entropy
production'', \ie the {\it time reversed processes occur with equal
likelyhood as the direct processes if conditioned to the opposite
entropy creation}. This is made clearer by rewriting the Eq.(\ref{e4.3})
in terms of probabilities {\it conditioned on a preassigned value
of $p$}; in fact up to $e^{O(1)}$ it becomes, \cite{Ga00}, for $|p|<p^*$:

\be\eqalign{
&\fra{P_\m(|F_j(S_tx)-\f_j(t)|_{j=1,\ldots,n}<\e,\, |\, p)}
{P_\m(|F_j(S_tx)\mp\f_j(-t)|_{j=1,\ldots,n}<\e, \, |\, -p)}=1\cr
}\label{e4.4}\ee

\*
\def\SEC{Consequences and comments}
\section{Consequences and comments}\iniz

\0(i) In {\it stationary states} of reversible dynamics heat
exchanges are constrained by (as remarked by Bonetto, \cite[Eq.(9.10.4)]{Ga00}),

\be\media {e^{-\ig_0^\t \sum_j\fra{Q_j(t)}{k_BT_j}\,dt}}=1,\label{e5.1}\ee
in the sense that $\fra1\t\log \media{\cdot}\tende{\t\to\io}0$.
Not to be confused with the formulae of \cite{BK81a} (and the
later developments) dealing with properties either of equilibrium
distributions or of distributions with density in phase space.
\*

\0(ii) It should not be thought that $p^*$ is proportional to the
maximum of the finite time averages of $\e(S_tx)$. The value of $p^*$
is the maximum value of $p$ observable {\it with a probability which
does not tend to zero faster than exponentially} as time tends to
$\infty$, see \cite{Ga95b}. This is analogous to the fact that in a
hard sphere gas the close packing density is not the maximum of the
density observable in finite volume.  \*

\0(iii) it has been claimed that the CH is not necessary to prove FT:
this is of course obvious. However some nontrivial assumption is
necessary: the CH is a simple general property that captures the
essential role of chaos, just as the harmonic oscillators systems
capture the essence of the ordered motions.  \*

\0(iv) The timed observations are closer to the physical applications
than the observations in continuous time but it is, at least
mathematically, interesting that the FT can be extended to continuous
time observations, \cite{Ge98}. In physical applications, however,
there may be an essential difference between the continuous version
and the discrete one because sometimes the interaction potentials are
modeled by forces which diverge at contact (e.g. when the interaction
is of Lennard-Jones type) or in some special configurations. Then one
cannot suppose that the system is Anosov because the spurious term
$\dot W$ in the phase space contraction, Eq.(\ref{e2.4}), can become
large with a probability that is ``just exponentially small'': and
this will affect the fluctuation relation, \cite{BGGZ05}. The problem
can be avoided by using timed observations: provided care is adopted
in the choice of the timing events. One simply has to choose them so
that the Poincar\'e section $\Si$ does not contain the singular
conficurations. In this way the contribution from the spurious terms,
which has the form $\fra1\t(W(\t)-W(0))$, Eq.(\ref{e2.7}), will tend
to zero as $t\to\io$ and will be eliminated from the statistics of the
entropy because $W$ will be bounded at the times $0$ and $\t$ where it
needs to be computed, \cite{CV03a,BGGZ05}.  \*

\0(v) In the checks of the fluctuation relation it is necessarily
true that the time $\t$ has to be kept finite: looking at the proof of
the FT, \cite{GC95,Ga95b}, the problem of the finite $\t$ corrections,
needed because the FT deals with an asymptotic property as
$\t\to\infty$, can be attacked and quantitatively studied, at least in
some cases, \cite{GZG05}, by following ideas employed to deal with
``finite size effects'' in statistical mechanics.
\*

\0(vi) The extended form of the FT, Eq.(\ref{e4.2}), has been used to
 show that in the limit of $0$ forcing the FT reduces to the ordinary
 fluctuation dissipation theorem, thus implying the Green-Kubo
 relations and Onsager reciprocity in reversible systems satisfying
 the CH, \cite{Ga96a}. However assuming time reversibility only at
 $0$-forcing and the CH it is sufficient to obtain the
 fluctuation-dissipation theorem, \cite{GR97}.  \* 

\0(vii) It has been claimed in the literature that the FT is a
 consequence of an analogous property of the equilibrium
 distributions: this is an erroneous claim, see \cite{CG99} for a
 counterexample (which can be easily extended to cover even cases of
 very chaotic systems (a remark by F.Bonetto)). It is not possible to
 infer a property valid on the zero volume attractor from a property
 checked outside it.  \*

\0(viii) It has been claimed that $\s_+>0$ is not necessary in the
proof of FT: this is also not correct. It is essential not only
because it appears in the denominator of the very definition of $p$
but because positivity is used in the proof, see \cite{Ga95b}.  The
error might be explained because the relation is written as a property
of the not normalized $A=p\s_+$ {\it without conditions on the size of
$A$}: which is very misleading because it deals with a quantity which
could be $0$ if $\s_+=0$. It is physically obvious that the relation
FT holds for $p$ in the domain of definition of $\z$ which is
certainly finite under the CH, \cite{Ga95b}.  \*

\0(ix) Since the chaotic hypothesis is never strictly speaking
realized one refers to Eq.(\ref{e4.1}),(\ref{e4.2}) as a {\it
fluctuation relation} (FR), and its test is a test of the chaotic
hypothesis, in analogy with the tests of the ergodic hypothesis. So
far there have been several studies of the FR via simulations. However
there are only preliminary experimental results in experiments
designed to check it in cases in which the system is not clearly
modeled by equations on which a complete theory is also possible,
\cite{BCG08}. A common feature to the attempts made so far to test the
FR is that the function $\z(p)$ turns out to be not convex: a nice
discussion of one of the reasons for this phenomenon can be found in
\cite{Za07}.  \*

\0(x) Perhaps the deepest consequence of the CH is the possibility of
a precise theory of coarse graining: see \cite{Ga08} for a heuristic
discussion from a Physicist viewpoint. The view stems out of the proof
in \cite{GC95,Ga95b} of the FT and explains it, see also
\cite{Ga95a}. Furthermore the precise formulation of coarse graining
leads to a discussion of the possibility of extending the notion of
entropy to systems in steady non equilibrium, \cite{Ga01}. It also
leads to an analysis of the irreversibility of processes and to a
quantitative evaluation of their ``degree of ireversibility'',
\cite{Ga06}.  \*

\0(xi) Reversibility is a delicate point: it might seem that it makes
any check of the FR impossible, except in simulations. This is
discussed in several places in the literature, see \cite{BGGZ06}. 
 \*

\0(xii) The CH is related, as mentioned, to the theory of
turbulence. Conversely the analysis of the CH anf the fluctuation
theorems has implications on the theory of turbulence, \cite{Ga06d,
Ga08,Ga02}.
\*

\0(xiii) The theory can be extended to quantum systems, \cite{Ga08},
modeled by finite thermostats. Considering the system in Fig.1 let $H$
be the operator on $L_2(\CC_0^{3N_0})$, of symmetric or antisymmetric
wave functions $\Ps$, $H= -\fra{\hbar^2}2\D_{\V X_0}+ U_0(\V
X_0)+\sum_{j>0}\big(U_{0j}(\V X_0,\V X_j)+U_j(\V X_j)+K_j\big)$,
parameterized by the configurations $(\V X_j,\dot{\V X}_j)_{j>0}$ of
the particles in the thermostats (here $K_j\defi\fra{m}2\dot{\V
X}_j^2$) and consider the {\it classical} dynamical system on
$\big(\Ps,(\{\V X_j\},$ $\{\V{{\dot X}}_j\})_{j>0}\big)$:

\be\eqalign{
-i\hbar & {\dot\Ps(\V X_0)}=\,(H\Ps)(\V X_0),\cr\noalign{\vskip3mm}
& \V{{\ddot X}}_j=-\Big(\dpr_j U_j(\V X_j)+
\media{\dpr_j U_j(\V X_0,\V X_j)}_\Ps\Big)-\a_j \V{{\dot X}}_j\kern4mm
j>0\cr}\label{e5.2}\ee
where the multipliers $\a_j$ are such to constrain the classical
thermostats to have a constant kinetic energy $K_j=\fra32 N_j k_B T_j$

\be\eqalign{
& \a_j\defi\fra{\media{W_j}_\Ps-\dot U_j}{2 K_j}, \qquad
W_j\defi -\V{{\dot X}}_j\cdot \V\dpr_j U_{0j}(\V X_0,\V X_j)\cr
&\s(\V X,\dot{\V X})\=\e(\V X,\dot{\V X})+
{\dot W(\V X)}\defi\sum_{j>0}\fra{Q_j}{k_B T_j}+\dot W\cr}\label{e5.3}\ee
which is time reversal symmetric if $I(\Ps,\V X,\dot{\V X})\defi
(\lis\Ps, \V X,-\dot{\V X})$. If the CH is assumed the 
FT is expected to hold for this model, hence for the entropy creation
rate,  $\e$. See for more details \cite{Ga08}.
\*

\0{\it Acknolwdgements:} {\small This paper was the text prepared for my talk
at StatPhys23, given on the occasion of the Boltzmann medal ceremony.
I wish to thank, in this occasion, {\it Daniela} (wife) and {\it
Barbara} (daughter), my teachers to whom I owe what I understand and
the way to look at Physics {\it Prof. Bruno Touschek, Prof. David
Ruelle, Prof. Joel Lebowitz, Prof. EGD Cohen}, my collaborators: from
all I learnt far more than it might have appeared at times; in
particular {\it Salvador Miracle-Sol\'e, Giuseppe Benfatto} and more
recently {\it Vieri Mastropietro, Guido Gentile, Federico
Bonetto, Alessandro Giuliani}.  It is a great honor to have been
selected together with {\it Kurt Binder}. The actual lecture was
somewhat different, due to time restrictions, and the transparencies can
be found in the conference web-site, {\tt http://www.statphys23.org}.}

\def\SEC{References}\iniz
\label{secRef}
{\small

\bibliographystyle{unsrt}
}
\end{document}